\documentclass[aip,pop,amsmath,amssymb,reprint,groupedaddress]{revtex4-1}

\usepackage{bm,times}

\newcommand{\beq}{\begin{equation}}
\newcommand{\eeq}{\end{equation}}
\newcommand{\bea}{\begin{eqnarray}}
\newcommand{\eea}{\end{eqnarray}}
\newcommand{\req}[1]{Eq.\ (\ref{#1})}
\newcommand{\kB}{k_\mathrm{B}}

\begin{document}

\title{Comment on ``On the ionization equilibrium of hot hydrogen
plasma\\[0pt] and thermodynamic consistency of formulating finite partition
functions'' \\[0pt] [Phys.\ Plasmas 17, 062701 (2010)]\\[1ex]
{\large \textit{Published:}
Phys. Plasmas \textbf{17}, 124705 (2010);
\textit{supplemented with two appendices}\\[0ex]~}}

\author{A. Y. Potekhin}
    \affiliation{Ioffe Physical-Technical Institute,
     194021 St.\ Petersburg, Russia}
\date{March 8, 2011}

\begin{abstract}

Zaghloul [Phys.\ Plasmas \textbf{17}, 062701 (2010)] reconsiders the  occupation probability
formalism  in plasma thermodynamics and claims inconsistencies in
previous models. I show that the origin of this incorrect claim
is an omission of the configurational factor from the partition
function. This arXiv version is supplemented with two appendices,
where I add remarks and comments on two more recent publications
of the same author on the same subject: on his response to this
Comment [Phys.\ Plasmas \textbf{17}, 124705 (2010)] and on his
criticism towards the Hummer and Mihalas's (1988)
formalism [Phys.\ Plasmas \textbf{17}, 122903 (2010)].

\end{abstract}


\maketitle

In a recent paper, Zaghloul\cite{Zag} revised 
the occupation probability formalism routinely applied for
quenching divergencies in frames of the chemical
picture of plasmas.\cite{HM88,P96} Following Ref.~\citenum{P96},
he considers a plasma composed of
protons, electrons, and H atoms and writes separate
expressions for the contributions of these subsystems into
the free energy:
$F_e$, $F_p$, and $F_\mathrm{H}$, respectively.
The atomic contribution is written 
in the form
\beq
   F_\mathrm{H}=N_\mathrm{H} \kB T
   \left[\ln\bigg(\frac{N_\mathrm{H}\lambda_\mathrm{H}^3
   }{
   VQ_{\mathrm{int},\mathrm{H}}}\bigg)-1\right],
\label{F2}
\eeq
where  $\kB$ is Boltzmann constant, $T$ temperature,
$N_\mathrm{H}$ the total number of atoms in all quantum states,
$\lambda_\mathrm{H} = 
(2\pi\hbar^2/m \kB T)^{1/2}$ the thermal wavelength of an atom,
$m$ is the atomic mass,
and
$Q_{\mathrm{int},\mathrm{H}}$ is the internal partition function. 
The author
fails to notice that \req{F2} is valid only for a
Boltzmann gas of \emph{noninteracting} particles (e.g.,
Ref.~\citenum{LaLi}, \S\S\,41, 42).
In general, instead of \req{F2} one should start from
the expression $F=-\kB T\,\mathrm{Tr}\, e^{-\hat{H}}$, where
$\hat{H}$ is the total Hamiltonian of the system (e.g.,
Ref.~\citenum{LaLi}, \S\,31).
Assuming that (i) the motion of particles is quasi-classical, (ii) the
kinetic and potential energies in $\hat{H}$ are uncoupled, 
(iii) interactions between plasma particles appear
in $\hat{H}$ as an additive
potential function,
one has\cite{GHR69,FGV77}
$
  F=-\kB T\ln Z=-\kB T\ln
   (Z_\mathrm{trans}Z_\mathrm{int}Z_\mathrm{conf})=
   F_\mathrm{trans}+F_\mathrm{int}+F_\mathrm{conf},
$
where the first two terms correspond to the translational and
internal degrees of freedom and 
the third one
 takes
 into account interactions between all plasma particles
 (in general, not only those between neutral
 atoms).
In the case of H atoms,
 $
 \ln Z_\mathrm{trans,H}=-F_\mathrm{trans,H}/\kB T=
 N_\mathrm{H} \ln(eV/N_\mathrm{H}\lambda_\mathrm{H}^3)$.
Having defined
$Q_{\mathrm{conf}}=Z_{\mathrm{conf}}^{1/N_\mathrm{H}}$
and
$Q_{\mathrm{int}}=Z_\mathrm{int}^{1/N_\mathrm{H}}$,
one can write
\beq
   F_\mathrm{H}=N_\mathrm{H} \kB  T
   \left[\ln\bigg(\frac{N_\mathrm{H}\lambda_\mathrm{H}^3
   }{   VQ_{\mathrm{int}}Q_{\mathrm{conf}}}\bigg)-1\right].
\label{F3}
\eeq
In general, \req{F3} cannot be reduced to \req{F2}.
Moreover, since level populations depend on interactions in the plasma,
$Q_{\mathrm{int}}$ in \req{F3} may differ from 
$Q_{\mathrm{int},\mathrm{H}}$ for the ideal Boltzmann gas in \req{F2}
(it is well known\cite{GHR69,HM88} that $Q_{\mathrm{int},\mathrm{H}}$
needs a cutoff to avoid divergency
due to the infinite number of shallow Rydberg states).
Conversely, $Q_\mathrm{conf}$ depends on internal level populations,
because interaction forces between atoms depend on their
excitation states. Thus, $F_\mathrm{int}$ and $F_\mathrm{conf}$
are not independent, and the definition of $F_\mathrm{int}$
is not obvious.

The free energy minimization method assumes that
$F$ is expressed explicitly
through numbers of particles of different kinds
and minimized with respect to these numbers at constant volume $V$.
In our case, $F=F(\{N_\kappa\},N_e,N_p)$,
where $N_\kappa$ are numbers of atoms on quantum levels $\kappa$.
Let us calculate
$F_\mathrm{id} \equiv F_\mathrm{trans}+F_\mathrm{int}$
using relation\cite{LaLi} $F=\bar{E}-TS$, where $\bar{E}$
is the mean energy and $S$
is the entropy. 
Assuming that the plasma is uniform in space, and
motion of atoms
is classical with distribution density $\mathcal{F}_\kappa(\bm{p})$
over momenta $\bm{p}$, 
the contribution of $N_\kappa$ atoms
to $\bar{E}$
is 
$
  N_\kappa \int d^3p \,
  \mathcal{F}_\kappa(\bm{p})\,\epsilon_\kappa(\bm{p}),
$
where $\epsilon_\kappa(\bm{p})$ is the total (kinetic minus binding)
atomic energy, while the entropy contribution is
$
  - \kB  N_\kappa \int d^3p \,
  \mathcal{F}_\kappa(\bm{p})
  \,\ln[\mathcal{F}_\kappa(\bm{p})\,(2\pi\hbar)^3 N_\kappa/g_\kappa eV)]
$
where $g_\kappa$ is
 quantum degeneracy of
level $\kappa$.
Let us consider the case where
$\epsilon_\kappa(\bm{p})
=p^2/2m - \chi_\kappa$ 
and binding energies $\chi_\kappa$
do not depend on $\bm{p}$
(a more general case has been studied in Ref.~\citenum{PCS}).
Then $\mathcal{F}_\kappa(\bm{p})=
(\lambda_\mathrm{H}/2\pi\hbar)^3\,e^{-p^2/2m\kB T}.
$
After integration and adding 
the translational contribution of $N_p$ classical protons
and the contribution of electron gas $F_\mathrm{id,e}$,
one obtains
\bea
   F_\mathrm{id} &=& \kB  T \sum_\kappa N_\kappa  
    \ln(e^{-\chi_\kappa/\kB  T -1}N_\kappa\lambda_\mathrm{H}^3/g_\kappa V)
\nonumber\\&&
+ \kB  T N_p\left[\ln(N_p\lambda_p^3/V)-1 \right]
 +F_\mathrm{id,e},
\label{Fid}
\eea
where $\lambda_p$ is the proton thermal wavelength.
For brevity we shall approximate $\lambda_p=\lambda_\mathrm{H}$.
The minimum of $F=F_\mathrm{id}+F_\mathrm{conf}$ 
under the stoichiometric 
constraints with respect to dissociation/recombination
reactions
H${}\leftrightarrows e+p$
requires 
\beq
   \frac{ \partial F}{\partial N_\kappa} = 
   \frac{\partial F}{\partial N_p} + 
    \frac{\partial F}{\partial N_e} .
\label{min}
\eeq
This gives, with account of \req{Fid},
\beq
   \ln\left(\frac{N_\kappa/g_\kappa}{N_p}\right) = 
   \frac{\chi_\kappa+\mu_e}{\kB  T} + 
\frac{\partial f}{\partial N_p} + 
    \frac{\partial f}{\partial N_e} - 
    \frac{\partial f}{\partial N_\kappa},
\label{NkNp}
\eeq
where $\mu_e=\partial F_\mathrm{id,e}/\partial N_e$ and 
$f=F_\mathrm{conf}/\kB  T$.

An occupation probability $w_\kappa$ is conventionally
 defined\cite{HM88}
as the probability of finding the atom in state $\kappa$ relative
to finding it in a similar ensemble of noninteracting ions.
In our case this means that
$N_\kappa \propto w_\kappa g_\kappa e^{\chi_\kappa/\kB T}$.
Therefore,
according to \req{NkNp}, 
$\ln w_\kappa = - \partial f / \partial N_\kappa + C_\mathrm{H}$,
where $C_\mathrm{H}$ does not depend on $N_\kappa$.
Thus one can write
\beq
 \frac{N_\kappa}{N_\mathrm{H}}=\frac{w_\kappa g_\kappa
e^{\chi_\kappa/\kB T}}{Q_{\mathrm{int},\mathrm{H},w}},
\label{Nk}
\eeq
where
\beq
 Q_{\mathrm{int},\mathrm{H},w} =
    \sum_\kappa g_\kappa w_\kappa e^{\chi_\kappa/\kB  T}.
\label{Zw}
\eeq
Note that number fractions $N_\kappa/N_\mathrm{H}$ do not depend on $C_\mathrm{H}$.
Hummer \& Mihalas\cite{HM88} set $C_\mathrm{H}=0$.
However, an additional requirement that 
the equation of ionization equilibrium for nondegenerate plasma
has the form of Saha equation
multiplied by $w_\kappa$
[$N_\kappa \propto N_p N_e w_\kappa e^{\chi_\kappa/\kB T}$; see
Eq.~(17) of Ref.~\citenum{P96}]
leads to
\beq
   \ln w_\kappa = \frac{\partial f }{ {\partial N_p}}
     +\frac{ \partial f }{ {\partial N_e}}
     - \frac{\partial f }{ {\partial N_\kappa}} + C_{\mathrm{H},e,p},
\label{w}
\eeq
where $C_{\mathrm{H},e,p}$ is independent of 
$N_\kappa$, $N_e$, and $N_p$. 
Given the constraints $N_\mathrm{H}=\sum_\kappa N_\kappa$
and $N_\mathrm{H}+N_p=$constant,
it is easy to
see that $N_\kappa$
do not depend on the choice of $C_{\mathrm{H},e,p}$.
We set\cite{P96,PCS} $C_{\mathrm{H},e,p}=0$
 (then obviously $C_\mathrm{H}=\partial f / {\partial N_p}
     + \partial f / {\partial N_e}$).

Substitution of (\ref{Nk}) into (\ref{Fid}) gives
\bea
F_\mathrm{id} &=& \kB  T N_\mathrm{H} \left[\ln({N_\mathrm{H}\lambda_\mathrm{H}^3}/{V})-1\right]
\nonumber\\&&
+\kB  T N_p \left[ \ln({N_p\lambda_p^3}/{V})-1 \right]
+ F_\mathrm{id,e} +F_\mathrm{int},
\eea
where
\beq
   F_\mathrm{int} = -\kB  T N_\mathrm{H} \ln Q_{\mathrm{int},\mathrm{H},w}
   + \kB  T \sum_\kappa N_\kappa\ln w_\kappa.
\label{Fint}
\eeq

Note that $Q_{\mathrm{int},\mathrm{H},w}$
appears in (\ref{Nk})
merely as a normalization
constant, and the occupation probabilities $w_\kappa$ 
are auxiliary quantities, 
defined from the condition of the minimum of the total
free energy according to \req{w}.

Zaghloul\cite{Zag} follows another route.
He replaces $Q_{\mathrm{int},\mathrm{H}}$ 
by $Q_{\mathrm{int},\mathrm{H},w}$
in
Eq.~(\ref{F2}),
leaving the meaning of quantities $w_\kappa$ undefined,
and assumes that this replacement is a
way of accounting for the nonideality
effects, \emph{alternative} to the introduction of
$F_\mathrm{conf}$ (as he explicitly writes 
and exposes in his Eq.~26).
This implies that the product
$Q_{\mathrm{int}}Q_{\mathrm{conf}}$ in \req{F3}
can be represented as a single sum (\ref{Zw}).
In general, it cannot. Furthermore,
this assumption leads to an additional
restriction on $w_\kappa$ (Eq.~32 of Ref.~\citenum{Zag}),
which may not necessarily be fulfilled in a real plasma.

We should remark that the expression for the free energy
can be written through $w_\kappa$ without
$F_\mathrm{conf}$ in the ``low-excitation approximation''
of Hummer \& Mihalas,\cite{HM88} who write it in the form
$f-\sum_\kappa N_\kappa\partial f/\partial N_\kappa=0$.
Taking into account that they consider the case where $C_\mathrm{H}=0$,
this approximation can also be written as
\beq
  F_\mathrm{conf} + \kB T \sum_\kappa N_\kappa \ln w_\kappa = 0.
\label{low-ex}
\eeq
The latter form is more general.
When condition (\ref{low-ex}) is satisfied, 
the second term in \req{Fint} annihilates with the
configurational part $F_\mathrm{conf}$ 
of the total Helmholtz free energy
$F=F_\mathrm{trans}+F_\mathrm{int}+F_\mathrm{conf}$.

The low-excitation approximation has serious shortcomings
(see discussion in \S\,IIId of Ref.~\citenum{HM88}).
One can explicitly show that it is violated
in some thermodynamic models commonly used in literature
(for instance, the hard-sphere
model\cite{HM88}).
For these reasons,
approximation (\ref{low-ex}) is used rather rarely.
In particular, it was not employed in
Refs.~\citenum{P96} and \citenum{PCS}. Without this approximation,
however, $F=F_\mathrm{id}+F_\mathrm{conf}$
does not reduce to an expression containing only $w_\kappa$
without $F_\mathrm{conf}$,
as required in Ref.~\citenum{Zag}.

In short, the conclusions in Ref.~\citenum{Zag} originate
from a trivial error: the author arbitrarily removes from
the partition function the configurational factor that is
responsible for interactions between plasma particles, however
assumes the significance of such interactions by allowing
occupation probabilities to differ from unity. 
The controversies in Ref.~\citenum{Zag} result
from this basic omission and not from the alleged
inconsistencies of the previous models.

\vspace*{3ex}
This work was partially supported by
Rosnauka Grant 
NSh-3769.2010.2
and RFBR Grant 08-02-00837. 

\appendix
\section{\label{sec:A}Comment on Zaghloul's response [Phys.\ Plasmas
\textbf{17}, 124705 (2010)]}

After the publication of this Comment\cite{commenz},
Zaghloul responded\cite{ZagResp} with
twelve items of arguments, one of which is subdivided into four
subitems. Here I briefly comment on them, item by item.

(i) \textit{``Equation (2) can always be reduced to Eq. (1) by defining
$Q_\mathrm{int} Q_\mathrm{conf} = Q_{\mathrm{int,H},w}$''} --- but in this case
$Q_{\mathrm{int,H},w}$ will not have the required form
of a partition function with weights $w_\kappa$, \req{Zw}.

(ii) \textit{``Following Eq.~(2), the author writes: `Moreover, since level populations depend on interactions in the plasma,
$Q_{\mathrm{int}}$ in \req{F3} may differ from 
$Q_{\mathrm{int},\mathrm{H}}$ for the ideal Boltzmann gas in \req{F2}
(it is well known\cite{GHR69,HM88} that $Q_{\mathrm{int},\mathrm{H}}$
needs a cutoff to avoid divergency
due to the infinite number of shallow Rydberg states).
Conversely, $Q_{\mathrm{conf}}$ depends on internal level populations,
because interaction forces between atoms depend on their
excitation states. Thus, $F_\mathrm{int}$ and $F_\mathrm{conf}$
are not independent, and the definition of $F_\mathrm{int}$
is not obvious.' Clearly, this is the main message of our
paper.''} --- Here Zaghloul agrees that $F_\mathrm{int}$ and
$F_\mathrm{conf}$ are not independent. However, next he writes:
\textit{``according to factorizability of the PF adopted in
[the Comment], $Q_\mathrm{int}$ in Eq.~(2) should be $Q_\mathrm{int,H}$
as in Eq.~(1).''} --- This is not logical.
Just the opposite,
since $F_\mathrm{int}$ and $F_\mathrm{conf}$ are not independent, the presence of
nonideality (i.e., $F_\mathrm{conf}$) modifies level populations and thus
stipulates the modification of $Q_\mathrm{int,H}$ which transforms into
$Q_\mathrm{int,H,w}$.

(iii) Having cited \req{Nk}, Zaghloul then writes: \textit{``the
above definition of $w_\kappa$ can be written as 
$w_\kappa=(N_{\kappa,w}/N_\mathrm{H})/(N_\kappa/N_\mathrm{H})
 = N_{\kappa,w}/N_\kappa$.''} --- This equation is mathematically
 wrong, as its left-hand side does not agree with \req{Nk}.
Therefore, the subsequent discussion in this item of
is incorrect as well.
 
(iv) \textit{``The substitution of Eq.~(6) into Eq.~(3) [\ldots]
represents an inconsistency [\ldots], where Eq.~(3), which
embodies uncoupling between the translational and internal
energies, is derived using a Boltzmann distribution of the
excited states while Eq.~(6) represents a different distribution
(real distribution [\ldots])''} --- in fact, however, Eq.(3) is
derived using the real distribution (which has to be determined);
it is not the ideal-gas Boltzmann distribution.

(v) Discussing the last term on the right-hand side of
\req{Fint}, Zaghloul writes: \textit{``The negative of this term,
namely, $-\kB T\sum_\kappa N_\kappa\ln w_\kappa$, was interpreted
[\ldots] as a contribution to the ideal-gas part of the entropy
due to correction $w_\kappa$ to the probability that the
$\kappa$th state is occupied. However, this is physically
incorrect since $w_\kappa$ is less than unity by definition. As a
result, the contribution stated above is positive which means
[\ldots] that using the occupation probabilities $w_\kappa$,
which diminishes levels' degeneracies, will increase the entropy,
which is in direct contradiction with the fundamental statistical
interpretation of the entropy!.''} --- First, 
in frames of the formalism under
consideration\cite{HM88,P96}
(where, for simplicity, quantum-mechanical
bound level shifts and broadening
in the plasma environment are neglected),
using the
occupation probabilities $w_\kappa$ does not ``diminish the
levels' degeneracies'': quantum-mechanical degeneracies remain
the same, $g_\kappa$. In this approach,
the occupation probabilities are
additional factors, which have a different physical meaning.
Second, the factor $w_\kappa$, as defined in \req{w}, is the
ratio of nonideal to ideal occupation numbers, which is not
necessarily less than unity (see discussion in
Ref.~\citenum{P96}). Third, the increase of the entropy with
deviation of the true distribution of level populations from the
Boltzmann distribution is not in ``contradiction,'' but in full
accord with fundamental principles of statistical thermodynamics.
Indeed, since the Boltzmann distribution provides the minimum
entropy (without interactions), then the use of $w_\kappa$ must
increase the part of the  entropy related to $F_\mathrm{int}$.

(vi) This item attempts to disprove \req{w} and
consists of four subitems (\textit{``several mistakes in the
derivation of this expression need to be unveiled here''}, writes
Zaghloul). They are commented in order below.

(vi-1) Equation (\ref{w}) is \textit{``based on the above-quoted
definition of $w_\kappa$. We have shown above that this
definition is inaccurate.''} --- Naturally, the equation is based
on the definition, there is no contradiction. As concerns the
alleged inaccuracy of this definition ``shown above,'' this
statement arises from Zaghloul's confusions listed (as well) above.

(vi-2) Equation (\ref{Nk}) \textit{``is in every respect
equivalent to \req{NkNp}. Since no efforts were devoted in the
derivation of \req{NkNp} to assure convergence of the IPF, then
there is no guarantee for the convergence of the IPF''} --- this
statement is true, but there is no contradiction, nor
inconsistency. Equations (\ref{NkNp}) and (\ref{Nk}) are quite
general, while the convergence depends on the existence of
$F_\mathrm{conf}$ and on its form.

(vi-3) \textit{``no efforts were made to assure that
$N_\kappa/N_\mathrm{H}$ satisfies the normalization condition
[\req{Zw}]''} --- this point is simply wrong, because,
given \req{Nk}, the normalization
condition $\sum_\kappa N_\kappa = N_\mathrm{H}$
is equivalent to definition (\ref{Zw}).

(vi-4) \textit{``Most importantly, we have shown
\underline{mathematically}
[\ldots] that for occupational probabilities similar to those
given by \req{w} [\ldots] the condition for the separability
of the configurational component [\ldots]
implies a linear dependence of
$F_\mathrm{conf}$ on the populations of individual excited
states''} --- in fact, this ``mathematical'' demonstration was
done in frames of the assumption that $F_\mathrm{conf}$ should be
dropped out as soon as $F_\mathrm{int}$ is modified. The
failure of this assumption is just in focus of my Comment.

(vii) This item consists of three paragraphs.
The first one contains two long quotations, one from the
Comment and another from the work by Hummer \&
Mihalas\cite{HM88}, both concerning the low-excitation
approximation. It is followed by two paragraphs of the criticism of the
Hummer \& Mihalas's work with a reference to another paper by
Zaghloul\cite{ZagHM}, especially devoted to this criticism. The
incorrectness of the latter paper\cite{ZagHM} is explained in
Appendix~\ref{sec:B} below.

(viii) This item consists of three paragraphs filled with
excerpts from the last part of the Comment, which deals
specifically with the
low-excitation approximation. Zaghloul claims it to be
\textit{``completely irrelevant as we do not use such an
approximation, and in fact we strongly criticized it''}, but
fails to note that his assumption of the
cancellation of  $F_\mathrm{conf}$ with the last term
in \req{Fint} is equivalent to
\req{low-ex}, that is equivalent to using the same low-excitation
approximation that he pretends to criticize.

The last paragraph of item (viii) ends up with the following
passage: \textit{``Axiomatically, the general case embodies the
special case as one of its possibilities. Now, is it possible
that the special case shows a restriction that the general case
does not show?''} --- Here, the special case is meant to be the
low-excitation approximation. Naturally, the special case always
is a restricted applicability relative to the general case,
therefore this rhetoric question is pointless.

(ix) This item consists in a reformulation of \req{Fid} (though
with some misprints), followed by the sentence: \textit{``It is
left to the reader to verify that the condition of the minimum''}
[of the free energy, \req{min}] \textit{``cannot lead to the
expression for $w_\kappa$ given by \req{w}.''} --- In fact, the
reader can easily see that \req{w} is the direct consequence of
the condition of the minimum of the free energy, as \req{w} is
derived from \req{Fid} and \req{min}, through \req{NkNp}, using
definitions (\ref{Nk}) and (\ref{Zw}).

(x) The author refers to his Ref.~\citenum{ZagHM} (to be dealt
with in Appendix~\ref{sec:B}), quotes the expression
$w=\exp(-F_\mathrm{conf}/N\kB T)$, and says: \textit{``the claim''}
[in the Comment] \textit{``that an error originates in our analysis from removing
the configurational component from the PF and assuming the
significance of such interactions by allowing occupation
probabilities to differ from unity, is groundless.}'' --- In fact,
this sentence is groundless itself, because the
Comment shows that Zaghloul's omission of $F_\mathrm{conf}$ from the
final expression of $F$
(after $F_\mathrm{int}$ has been modified by introducing
$w_\kappa$) is incorrect, in general, unless one additionally uses
the low-excitation approximation, \req{low-ex}.

(xi) The author states that summing over $\kappa$ in \req{w}
amounts to \textit{``the term commonly referred to in the
literature as a lowering of ionization energy''} --- 
which only confirms the consistency the results
exposed above with other well known results in the literature.
Zaghloul writes further: \textit{``Clearly, the inclusion of this
term cannot exclusively lead to a truncation of the IPF''}, which
is a mistake: it can lead to such truncation.

(xii) \textit{``For completeness, we would like to explain here
that \req{F2}, or more generally 
the relation $F_\mathrm{H} = -k_B T \ln Z$ is valid for
Boltzmann distribution for the total energy [\ldots] of the
system''} --- The latter relation, where $Z$ is the total partition
function of the system, including all ideal and nonideal
contributions, is a textbook fact. 
However, the general relation $F_\mathrm{H} = -k_B T \ln Z$ is
not equivalent to its particular case, \req{F2}.
In the Comment I called to the reader's attention,
that one should not substitute into this relation
$Z=Z_\mathrm{trans}Z_\mathrm{int}$, 
omitting $Z_\mathrm{conf}$, as
Zaghloul does.

\section{\label{sec:B}Comment on another similar paper
by Zaghloul [Phys.\ Plasmas
\textbf{17}, 122903 (2010)]}

A more recent paper\cite{ZagHM} of the same author is devoted to
the criticism of the formalism of Hummer \& Mihalas\cite{HM88}, which
lies in the basis of the Opacity Project (e.g.,
Ref.~\citenum{OP}, and references therein). There are several
mistakes in Ref.~\citenum{ZagHM}. One of them is that the author
mixes up the occupation probabilities $w_\kappa$ with quantum
statistical weights of the bound states $g_\kappa$,  writing
$w_i=g_i\omega_i\equiv\exp[-(\partial f/\partial N_i)/\kB T]$.
The statistical weight $g_i$ arises from counting multiply
degenerate quantum states, $g_i\geqslant1$, whereas 
$\exp[-(\partial f/\partial N_i)/\kB T]$ is most often less than $1$.
This mistake has no effect, for example, if one considers the
case of $g_i=1$. Another mistake is the statement that
\textit{``a monatomic perfect gas has only translational kinetic
energy (no means of storing energy except as kinetic energy).''}

However, there is a more serious fault that lies in the basis of
the article and hence disproves its main conclusions. 

The main focus of the criticism in this paper is the expression
$F=F_\mathrm{trans}+F_\mathrm{int}+F_\mathrm{conf}$. 
The author writes: 
\textit{``the factorizability of the partition function (or,
equivalently, the separability of the free energy components)
implies that various types of energies are independent of each
other''}. In fact, there is no such implication. He continues: 
\textit{``The separation of the configurational free energy
therefore indicates that it has no influence on the internal free
energy.''} In fact, it depends on the definition of the
$F_\mathrm{int}$ and $F_\mathrm{conf}$ terms, which is not
obvious in a nonideal system. If $F_\mathrm{int}$
is defined using the real occupation numbers (as done, e.g., in
Refs.~\citenum{P96,PCS} and in the
Comment above), then $F_\mathrm{conf}$ affects the
value of $F_\mathrm{int}$ at equilibrium through $N_\kappa$
values. In general, the minimum of
$F=F_\mathrm{int}+F_\mathrm{trans}+F_\mathrm{conf}$ realizes at a
different set of $\{N_\kappa\}$ than the minimum of
$F_\mathrm{int}+F_\mathrm{trans}$ alone.
In other words, the separability does not imply
independence. 

Furthermore, the author writes: \textit{``if the configurational
free energy (or interaction energy) has no influence on the
internal free energy, the expectation that including a separable
configurational component could lead to a truncation of the
internal partition function [\ldots] is physically and logically
questionable''} (in the e-print version [arXiv:1010.1102v1],
the word ``questionable'' is replaced by ``incorrect''),
\textit{``because they are independent of each other by
assumption''}. However, there is no such assumption in the cited
references. On the contrary, since both $F_\mathrm{int}$ and
$F_\mathrm{conf}$ are functions of the particle numbers
(including occupation numbers), they are interrelated.

\end{document}